\documentclass{article}
\usepackage{amssymb}

\input{tcilatex}
\begin{document}

\begin{center}
{\LARGE An Explicit Criterion for Existence of the Magnus Solution for a
Coupled Spin System under a Time-Dependent Radiofrequency Pulse{\large 
\footnote{%
This paper is the corrected version of the original manuscript (the year
2000) and its published paper (Xijia Miao, Phys. Lett. A 271 (2000)
296-302). There was not proof of the published paper. All the error
corrections, modifications, and comments are noted clearly in the footnotes
in this corrected paper. The corrected paper, the original manuscript and/or
the published paper appear in the website \textit{Arxiv.org} with the same
paper ID and different version.}}\bigskip }

{\large Xijia Miao}

{\large Laboratory of Magnetic Resonance and Atomic and Molecular Physics,
Wuhan Institute of Physics and Mathematics, The Chinese Academy of Sciences,
Wuhan 430071, P. R. China}

{\large (Current Email: mia\bigskip oxijia@yahoo.com)}

{\Large Abstract}
\end{center}

The explicit criterion is derived in detail for the convergence of the
Magnus expansion and the existence of the Magnus solution in the interaction
picture, i.e., the exponential propagator in the weakly coupled spin (I=1/2)
system S$_{n}$AMX... (n=1,2,3,...) in which only spin group $S$ is subjected
to a time-dependent shaped selective radiofrequency pulse. The derivation is
built on the same scheme as the Maricq's (J. Chem. Phys. 86 (1987) 5647). It
is shown that the criterion depends only upon amplitude of the
time-dependent field applied to the system, and the Magnus expansion
converges and the Magnus solution exists when the flip angle of a
non-negative-amplitude shaped RF pulse or a weak-amplitude shaped pulse is
smaller than $2\pi .$ The exponential propagator then can be decomposed into
a product of a series of elementary propagators and can be used to determine
time evolution of the spin system under the shaped pulse. The linear
differential equations to determine the unknown parameters in the Magnus
solution are obtained explicitly. An alternative propagator in an expansion
form for the Magnus solution is also proposed to describe the time evolution
when the criterion is not met. \newline
\footnote{%
This row of the original manuscript is already corrected here.}\newline
\newline
\newline

The Magnus expansion [1] plays an important role in determination for time
evolution of quantum systems under time-dependent Hamiltonian. The advantage
of the Magnus expansion over the standard quantum mechanical perturbation
theory is that Hermiticity is retained to each order in the expansion. The
Magnus expansion has an extensive application in nuclear magnetic resonance
(NMR) spectroscopy: the average Hamiltonian theory [2, 3] which is based on
the Magnus expansion has been used extensively in the design and analysis of
a variety of multipulse NMR experiments both in liquid and in solid states.
The Magnus expansion has also been used successfully to investigate the
response of coupled spin systems to a time-dependent radiofrequency (RF)
field, e.g., a shaped selective pulse [4, 5]. The Magnus solution, namely,
the continuous exponential solution to the time-dependent Schr\"{o}dinger
equation, has recently been used to characterize the behavior of a coupled
spin system under the time-dependent radiation damping field [6]. Although
the Magnus expansion and the Magnus solution are applied conveniently to a
variety of time-dependent systems, there are strict limits on the existence
of the solution and there is also a possible serious problem for the
expansion concerning its convergence [1, 7 - 9]. As pointed out by Magnus
[1] as long ago as in 1954, the Magnus solution $U(t)=\exp [-i\Omega (t)]$
exists only when the so-called Magnus criterion is satisfied: $|\lambda
_{i}(t)-\lambda _{j}(t)|\neq n2\pi ,$ where $\lambda _{i}(t)$ are the
eigenvalues of the operator $\Omega (t)$ and $n$ is a nonzero integer. This
criterion is an implicit one and has little value since $\lambda _{i}(t)$
are the unknown quantity under investigation. \textit{An explicit criterion%
\footnote{%
The original words are modified here. The original word(s) appears in the
original manuscript and its published paper (Xijia Miao, Phys. Lett. A 271
(2000) 296-302). Hereafter use italics to mark the modified word(s). }} is
therefore needed for the convenient application of the Magnus solution.
Maricq [8] showed that the operator $\Omega (t)$ exists for short
evolutional times and the Magnus expansion always converges in the
neighborhood of the time $t=0$. Maricq [9] further found that the
convergence for an isolated two-level system under a linear polarized field
can be improved: the Magnus expansion is convergent and hence the Magnus
solution exists in time interval $[0,$ $\tau ]$ if the following inequality
is satisfied 
\begin{equation}
\int_{0}^{\tau }|\omega (t^{\prime })|dt^{\prime }<2\pi  \tag{1}
\end{equation}%
where $\pm \omega (t)/2$ are the two eigenvalues of the time-dependent
Hamiltonian $H(t)$ of the system. The advantage for this criterion is that
it is dependent only on the eigenvalues of the Hamiltonian $H(t)$. However,
an explicit criterion may be far from the simple form of inequality (1) and
has not yet been found for a general time-dependent multilevel system [7 -
9].

A shaped selective pulse is an amplitude-modulation and/or phase-modulat-
ion RF field [4, 10 - 13]. The Magnus solution may provide a convenient
starting point for an exact description of time evolution of coupled spin
systems under a time-dependent shaped selective RF pulse. However, so far an
explicit \textit{criterion\footnote{%
The original word is modified here.}} for existence of the Magnus solution
and convergence of the Magnus expansion has not \textit{been\footnote{%
The original word is modified here.}} established for the general spin
systems. This general problem could not be settled simply, but in some
special cases it may be able to be solved without much difficulty. In this
communication the Maricq's criterion (1)\ for a single two-level system is
extended to the weakly coupled spin (I=1/2) systems S$_{n}$AMX...
(n=1,2,...), where only spin S is undergone a time-dependent RF field. It is
known that the properties of the Magnus expansion depend strongly on the
representation [7 - 9]. The present criterion is derived explicitly in the
interaction picture defined by spin Hamiltonian of the spin system in the
absence of the time-dependent RF field. In this representation the criterion
is dependent only upon the amplitude of the time-dependent RF field applied
to the system. Therefore, in practice this is quite convenient to use the
criterion.

\QTP{Body Math}
For simplification, consider the SAMX... weakly coupled spin (I=1/2) system
in which only spin S is subjected to a time-dependent RF field, e.g., a
shaped selective pulse. The contribution of the time-dependent RF field to
spin Hamiltonian of the system is given by%
\begin{equation}
H_{1}(t)=\omega _{1}(t)\{S_{x}\cos \phi (t)+S_{y}\sin \phi (t)\}  \tag{2}
\end{equation}%
where $\omega _{1}(t)$ and $\phi (t)$ are known as the amplitude and phase
of the time-dependent RF field, respectively. Then the total spin
Hamiltonian of the spin system in a strong static magnetic field can be
written in the rotating frame%
\begin{equation}
H(t)=H_{0}+H_{1}(t)  \tag{3}
\end{equation}%
where $H_{0}=H_{I}+\Omega _{I}S_{z};$ $H_{I}=\sum_{k}\Omega
_{k}I_{kz}+\sum_{k<l}\pi J_{kl}2I_{kz}I_{lz}$ and $\Omega _{I}=\Omega
_{s}+2\sum_{k}\pi J_{ks}I_{kz};$ the symbol $I$ denotes the spin A, M, X,
.... Obviously, both the \footnote{%
Here drops a redundant original word.} operators $H_{I}$ and $\Omega _{I}$
are members of the \textit{lo}ngitudinal \textit{m}agnetization and \textit{s%
}pin \textit{o}rder (\textit{LOMSO}) operator subspace of the Liouville
operator space of the spin subsystem AMX... of the system SAMX... [14]. They 
\footnote{%
Here drops a redundant original word.} commute with each other because any
two base operators of the \textit{LOMSO} operator subspace commute with each
other [14]. They also commute with the magnetization operators $\{S_{x}$, $%
S_{y}$, $S_{z}\}$ of spin $S$. The time-evolutional propagator corresponding
to the time-dependent Hamiltonian (3) can be written as 
\begin{equation}
U(t)=T\exp (-i\int_{0}^{t}H(t^{\prime })dt^{\prime })=\exp (-iH_{0}t)U_{I}(t)
\tag{4}
\end{equation}%
where $T$ is Dyson time-ordering operator and the propagator $U_{I}(t)$ in
the interaction representation is given by%
\begin{equation}
U_{I}(t)=T\exp (-i\int_{0}^{t}H_{i}(t^{\prime })dt^{\prime })  \tag{5}
\end{equation}%
with the time-dependent Hamiltonian in the interaction representation
defined by the unperturbed Hamiltonian $H_{0}$ of Eq. (3): 
\begin{equation}
H_{i}(t)=\exp (iH_{0}t)H_{1}(t)\exp (-iH_{0}t)  \tag{6}
\end{equation}%
By expanding the right-hand side of equation (6) according to the rotation
transformations [2] between any two product operators and the fact that $%
[H_{I},$ $\Omega _{I}]=0$, the Hamiltonian $H_{i}(t)$ is reduced to an
explicit expression%
\begin{equation}
H_{i}(t)=\omega _{1}(t)\boldsymbol{h(t).S}=\omega
_{1}(t)\{h_{x}(t)S_{x}+h_{y}(t)S_{y}+h_{z}(t)S_{z}\}  \tag{7}
\end{equation}%
where the operator $h_{x}(t)=\cos (-\Omega _{I}t+\phi (t)),$ $h_{y}(t)=\sin
(-\Omega _{I}t+\phi (t)),$ and $h_{z}(t)=0.$ Because $%
h_{x}(t)^{2}+h_{y}(t)^{2}+h_{z}(t)^{2}=E$ (unit operator), the operator $%
\boldsymbol{h(t)}=(h_{x}(t),h_{y}(t),h_{z}(t))$ is an operator vector with
the unit magnitude. Obviously, the operator $\boldsymbol{h(t)}$ belongs the 
\textit{LOMSO} operator subspace of the subsystem AMX.... It follows from
Eq. (7) that 
\begin{equation}
||H_{i}(t)||\leq |\omega _{1}(t)|.||\boldsymbol{h(t)||.||S||}  \tag{8}
\end{equation}%
Therefore, all eigenvalues of the Hamiltonian $H_{i}(t)$ are bounded by the
amplitude $|\omega _{1}(t)|$ in magnitude of the time-dependent RF field. In
a single-spin system $\boldsymbol{h(t)}$ is a vector of simple scalars and
in this case the criterion, that is, inequality (1), was obtained explicitly
by Maricq [9], but in the SAMX... coupled spin system $\boldsymbol{h(t)}$
can be taken as a vector of the operators $h_{p}(t)$ $(p=x,y,z)$ that all
commutes one with each other because the operators $h_{p}(t)$ are members of
the \textit{LOMSO} operator subspace. In the present more general case the
criterion of the Magnus expansion in the interaction picture has not been
obtained explicitly and it is just the purpose of the present work to extend
the criterion (1) to the S$_{n}$AMX... coupled spin systems.

\QTP{Body Math}
In order to determine exactly and analytically time evolution of the coupled
spin system under a time-dependent RF field one needs to solve the
time-dependent Schr\"{o}dinger equation or equivalent Liouville-von Neumann
equation. However, it is usually difficult to solve exactly the
time-dependent dynamical equations and therefore to determine exactly the
density operator and the response of the system to the time-dependent RF
field. A numerical calculation is a general routine to solve time-dependent
problems, but it is usually not much helpful for one to gain a deeper
insight into the physical essence for the effect of a time-dependent RF
field on the time evolution and for the effect of scalar coupling on the
response. The approximated treatments that usually employ the perturbation
theory or the average Hamiltonian theory are valuable, but there are some
limitations and disadvantages of these approximated methods and therefore,
an exact solution to the time-dependent problems is still attractive. One
strategy to determine exactly and analytically time evolution of the coupled
spin system under a time-dependent RF field is that the propagator $U_{I}(t)$
of Eq. (5) is written as an exponential operator without the Dyson
time-ordering operator T, as suggested by Magnus in early day [1]. Note that
the magnetization operators $\{E$, $S_{x}$, $S_{y}$, $S_{z}\}$ of the spin $%
S $ form the \textit{u(2)} Lie algebra subspace and the longitudinal
magnetization and spin order operators $\{E,$ $I_{kz},$ $2I_{kz}I_{lz},$ $%
4I_{kz}I_{lz}I_{mz},...\}$ of the subsystem AMX... form the \textit{LOMSO}
operator subspace [14] by including the unit operator $E$. Then the complete
operator set $\{E,$ $S_{p},$ $I_{kz},$ $2S_{p}I_{kz},$ $2I_{kz}I_{lz},$ $%
4S_{p}I_{kz}I_{lz},$ $...;$ $p=x,y,z\}$ is a closed operator algebra
subspace of the Liouville operator space of the spin system SAMX... and
actually, this subspace is the direct product space of the subspaces \textit{%
u(2)} and \textit{LOMSO}, that is, $G=u(2)\dbigotimes LOMSO$ [14].
Obviously, the Hamiltonian $H_{i}(t)$ of Eq. (7) is a member of the operator
subspace $G$ because the operators $h_{p}(t)$ $(p=x,y,z)$ in Eq. (7) are
members of the \textit{LOMSO} operator subspace. Then the propagator $%
U_{I}(t)$ of Eq. (5) corresponding to the time-dependent Hamiltonian $%
H_{i}(t)$ is also a member of the subspace $G$ [14] and may be conveniently
written as an exponential unitary operator: 
\begin{equation}
U_{I}(t)=\exp [-i\Omega (t)]  \tag{9}
\end{equation}%
with the Hermitian operator 
\begin{equation}
\Omega (t)=\Omega _{x}(t)S_{x}+\Omega _{y}(t)S_{y}+\Omega _{z}(t)S_{z} 
\tag{10}
\end{equation}%
where the Hermitian operator $\Omega (t)$ is an element of the operator
subspace $G$ and $\Omega _{p}(t)$ $(p=x,y,z)$ belong the \textit{LOMSO}
operator subspace. The exponential propagator of Eq. (9) is quite simple and
can be further decomposed into a sequence of elementary propagators [15, 16]%
\[
U_{I}(t)=\exp (-i\alpha (t)S_{z})\exp (-i\beta (t)S_{y})\exp (-i\hat{\Omega}%
(t)S_{z}) 
\]%
\begin{equation}
\times \exp (i\beta (t)S_{y})\exp (i\alpha (t)S_{z})  \tag{11}
\end{equation}%
where the operators $\alpha (t),$ $\beta (t),$ and $\hat{\Omega}(t)$ that
are also members of the \textit{LOMSO} operator subspace are related to the
operators $\Omega _{x}(t),$ $\Omega _{y}(t),$ and $\Omega _{z}(t)$ in Eq.
(10) by the following formulae:%
\begin{equation}
\Omega _{x}(t)\sin \alpha (t)=\Omega _{y}(t)\cos \alpha (t)  \tag{12a}
\end{equation}%
\begin{equation}
\Omega _{z}(t)\sin \beta (t)=[\Omega _{x}(t)\cos \alpha (t)+\Omega
_{y}(t)\sin \alpha (t)]\cos \beta (t)  \tag{12b}
\end{equation}%
\begin{equation}
\hat{\Omega}(t)=\Omega _{z}(t)\cos \beta (t)+[\Omega _{x}(t)\cos \alpha
(t)+\Omega _{y}(t)\sin \alpha (t)]\sin \beta (t)  \tag{12c}
\end{equation}

\QTP{Body Math}
The exponential propagator of Eq. (9) of the SAMX... spin system can be
extended to the coupled spin systems S$_{n}$AMX... (n=1,2,3,...) and in this
more general case the propagator can be written as\footnote{%
The incorrect expression (13) appearing in the original manuscript and its
published paper is 
\begin{equation}
U_{I}(t)=\dprod\limits_{k=1}^{n}\exp [-i\Omega _{x}(t)S_{kx}+\Omega
_{y}(t)S_{ky}+\Omega _{z}(t)S_{kz}]  \tag{13}
\end{equation}%
}%
\begin{equation}
U_{I}(t)=\dprod\limits_{k=1}^{n}\exp [-i(\Omega _{x}(t)S_{kx}+\Omega
_{y}(t)S_{ky}+\Omega _{z}(t)S_{kz})]  \tag{13}
\end{equation}

\QTP{Body Math}
With the help of the decomposed propagator of Eq. (11) one can determine
conveniently time evolution of the coupled spin system SAMX... under a
shaped pulse. However, there are two problems that need to be solved in
advance. One of which is that the existence of the exponential propagator $%
U_{I}(t)$ of Eq. (9) needs to be verified. This would be achieved by setting
up the convergence condition of the Magnus expansion of the operator $\Omega
(t)$ in Eq. (9) and hence the existence condition of the Magnus solution of
Eq. (9) in the interaction picture. Another is that the unknown parameters
in the decomposed propagator of Eq. (11) should be determined explicitly. It
was shown by Maricq [9] that the explicit condition (1) for the convergence
of the Magnus expansion in a single-spin system under a linear polarized
field is related only to the eigenvalues of the time-dependent Hamiltonian
of the system. However, it has not been clear whether the convergence of the
Magnus expansion in the coupled spin \textit{system\footnote{%
The original word is modified here.}} SAMX... is still dependent only upon
the eigenvalues of the time-dependent Hamiltonian of the system. From the
point of view of convenient application, the convergence had better depend
only on the eigenvalues of the Hamiltonian or even only upon the amplitude
and/or phase of the time-dependent RF field applied to the system in the
interaction frame. On the other hand, it is expected that the convergence of
the Magnus expansion in the interaction frame may depend only upon the
amplitude of the time-dependent RF field as all the eigenvalues of the
time-dependent Hamiltonian (6) in the interaction frame are bounded on by
the amplitude of the RF field in magnitude (see inequality (8)). Therefore,
in the following the convergence criterion is derived in the interaction
representation enabling it to depend only upon the amplitude of the RF
field. The derivation is based on the decomposed propagator of Eq. (11) and
the time-dependent Schr\"{o}dinger equation in the interaction frame:%
\begin{equation}
\frac{d}{dt}U_{I}(t)=-iH_{i}(t)U_{I}(t)\text{ \ }(\hslash =1)  \tag{14}
\end{equation}%
The decomposed propagator of Eq. (11) is first expanded as a linear
combination of the magnetization operators $S_{p}$ $(p=x,y,z)$%
\[
U_{I}(t)=\cos (\frac{1}{2}\hat{\Omega}(t))-2i[S_{z}\cos \beta (t)+S_{x}\cos
\alpha (t)\sin \beta (t) 
\]%
\begin{equation}
+S_{y}\sin \alpha (t)\sin \beta (t)]\sin (\frac{1}{2}\hat{\Omega}(t)) 
\tag{15}
\end{equation}%
Equation (15) is the expansion expression of the propagator $U_{I}(t)$ in
the operator subspace $G$. Then inserting both the Hamiltonian $H_{i}(t)$ of
Eq. (7) and the expansion-form propagator $U_{I}(t)$ of Eq. (15) into
equation (14) one obtains a set of equations entirely equivalent to the
time-dependent Schr\"{o}dinger equation (14)%
\[
\frac{d}{dt}\cos (\frac{1}{2}\hat{\Omega}(t))=-\frac{1}{2}\omega _{1}(t)\sin
(\frac{1}{2}\hat{\Omega}(t))\{h_{x}(t)\cos \alpha (t)\sin \beta (t) 
\]%
\begin{equation}
+h_{y}(t)\sin \alpha (t)\sin \beta (t)+h_{z}(t)\cos \beta (t)\}  \tag{16a}
\end{equation}%
\[
\frac{d}{dt}[\cos \alpha (t)\sin \beta (t)\sin (\frac{1}{2}\hat{\Omega}(t)),%
\text{ }\sin \alpha (t)\sin \beta (t)\sin (\frac{1}{2}\hat{\Omega}(t)),\text{
}\cos \beta (t)\sin (\frac{1}{2}\hat{\Omega}(t))] 
\]%
\[
=\frac{1}{2}\omega _{1}(t)\cos (\frac{1}{2}\hat{\Omega}(t))[h_{x}(t),\text{ }%
h_{y}(t),\text{ }h_{z}(t)] 
\]%
\[
+\frac{1}{2}\omega _{1}(t)\sin (\frac{1}{2}\hat{\Omega}(t))[-h_{z}(t)\sin
\alpha (t)\sin \beta (t)+h_{y}(t)\cos \beta (t), 
\]%
\[
h_{z}(t)\cos \alpha (t)\sin \beta (t)-h_{x}(t)\cos \beta (t), 
\]%
\begin{equation}
h_{x}(t)\sin \alpha (t)\sin \beta (t)-h_{y}(t)\cos \alpha (t)\sin \beta (t)]
\tag{16b}
\end{equation}%
Equations (16)\footnote{%
The incorrect equations (16) appearing in the original manuscript and its
published paper are%
\[
\frac{d}{dt}\cos (\frac{1}{2}\hat{\Omega}(t))=-\omega _{1}(t)\sin (\frac{1}{2%
}\hat{\Omega}(t))\{h_{x}(t)\cos \alpha (t)\sin \beta (t) 
\]%
\begin{equation}
+h_{y}(t)\sin \alpha (t)\sin \beta (t)+h_{z}(t)\cos \beta (t)\}  \tag{16a}
\end{equation}%
\[
\frac{d}{dt}[\cos \alpha (t)\sin \beta (t)\sin (\frac{1}{2}\hat{\Omega}(t)),%
\text{ }\sin \alpha (t)\sin \beta (t)\sin (\frac{1}{2}\hat{\Omega}(t)),\text{
}\cos \beta (t)\sin (\frac{1}{2}\hat{\Omega}(t))] 
\]%
\[
=\frac{1}{2}\omega _{1}(t)\cos (\frac{1}{2}\hat{\Omega}(t))[h_{x}(t),\text{ }%
h_{y}(t),\text{ }h_{z}(t)] 
\]%
\[
+\frac{1}{2}\omega _{1}(t)\sin (\frac{1}{2}\hat{\Omega}(t))[h_{z}(t)\sin
\alpha (t)\sin \beta (t)-h_{y}(t)\cos \beta (t), 
\]%
\[
h_{z}(t)\cos \alpha (t)\sin \beta (t)-h_{x}(t)\cos \beta (t), 
\]%
\begin{equation}
h_{x}(t)\sin \alpha (t)\sin \beta (t)-h_{y}(t)\cos \alpha (t)\sin \beta (t)]
\tag{16b}
\end{equation}%
} are a set of linear coupled differential equations with the operator
variables: $\cos (\frac{1}{2}\hat{\Omega}(t)),$ $\cos \alpha (t)\sin \beta
(t)\sin (\frac{1}{2}\hat{\Omega}(t)),$ $\sin \alpha (t)\sin \beta (t)\sin (%
\frac{1}{2}\hat{\Omega}(t)),\footnote{%
The incorrect operator variable appearing only in the published paper is $%
\sin \alpha (t)\sin \beta (t):\sin (\frac{1}{2}\hat{\Omega}(t))$}$ and $\cos
\beta (t)\sin (\frac{1}{2}\hat{\Omega}(t))$ of the \textit{LOMSO} operator
subspace. They can be used to determine generally the unknown parameters in
the decomposed propagator of Eq. (11). Also, from them one may extract the
criterion of convergence of the Magnus expansion in the interaction frame.
Because any two base operators of the \textit{LOMSO} subspace commute and
the operator $\hat{\Omega}(t)$ belongs the \textit{LOMSO} subspace one has%
\begin{equation}
\frac{d}{dt}\cos (\frac{1}{2}\hat{\Omega}(t))=-\frac{1}{2}\sin (\frac{1}{2}%
\hat{\Omega}(t))\frac{d}{dt}\hat{\Omega}(t)  \tag{17}
\end{equation}%
Noting equation (17), equation (16a) leaves one a simple form%
\[
\frac{d}{dt}\hat{\Omega}(t)=\omega _{1}(t)\{h_{x}(t)\cos \alpha (t)\sin
\beta (t) 
\]%
\begin{equation}
+h_{y}(t)\sin \alpha (t)\sin \beta (t)+h_{z}(t)\cos \beta (t)\}  \tag{18}
\end{equation}%
The equation (18) can be integrated conveniently because all the operators
in the equation belong the \textit{LOMSO} subspace. By employing the initial
condition $\hat{\Omega}(0)=0$ the integration is written as\footnote{%
The incorrect equation\ (19) appearing only in the published paper is%
\begin{equation}
\hat{\Omega}(t)=\int_{0}^{t}dt^{\prime }\{\omega _{1}\}(t^{\prime })%
\boldsymbol{h(t}^{\prime }\boldsymbol{).g(t}^{\prime }\boldsymbol{)} 
\tag{19}
\end{equation}%
}%
\begin{equation}
\hat{\Omega}(t)=\int_{0}^{t}dt^{\prime }\{\omega _{1}(t^{\prime })%
\boldsymbol{h(t}^{\prime }\boldsymbol{).g(t}^{\prime }\boldsymbol{)}\} 
\tag{19}
\end{equation}%
where the operator vector with unit magnitude $\boldsymbol{g(t)}=(\cos
\alpha (t)\sin \beta (t),$ $\sin \alpha (t)$ $\times \sin \beta (t),$ $\cos
\beta (t)).$

\QTP{Body Math}
Now the eigenvalues of the operator $\Omega (t)$ in the Magnus solution (9)
are readily calculated in the direct product basis of eigenbase $|i\rangle $
and $|s\rangle $ of single-spin angular momentum operators $I_{kz}$ and $%
S_{z}$ by diagonalizing the operator $\Omega (t)$ (see Eq. (11))%
\begin{equation}
\lambda _{i}^{s}(t)=\langle i|\langle \Psi _{s}|\Omega (t)|\Psi _{s}\rangle
|i\rangle =\langle i|\langle s|\hat{\Omega}(t)S_{z}|s\rangle |i\rangle
=m_{s}\langle i|\hat{\Omega}(t)|i\rangle  \tag{20}
\end{equation}%
where the magnetic quantum number $m_{s}=+1/2$ and $-1/2$ for $s=0$ and $1$,
respectively. Because of the unit magnitude of the operator vectors $%
\boldsymbol{h(t)}$ and $\boldsymbol{g(t)}$ it follows from Eq. (19) that
each eigenvalue of the operator $\hat{\Omega}(t)$ is bounded by%
\begin{equation}
|\langle i|\hat{\Omega}(t)|i\rangle |\leq \int_{0}^{t}dt^{\prime }|\omega
_{1}(t^{\prime })|  \tag{21}
\end{equation}%
This inequality (21) together with equation (20) can lead to directly the
desired result by the triangle inequality:%
\begin{equation}
|\lambda _{i}^{s}(t)-\lambda _{j}^{s^{\prime }}(t)|\leq |\lambda
_{i}^{s}(t)|+|\lambda _{j}^{s^{\prime }}(t)|\leq \int_{0}^{t}dt^{\prime
}|\omega _{1}(t^{\prime })|  \tag{22}
\end{equation}%
By comparing the inequality (22) with the Magnus criterion one obtains the
explicit criterion of convergence of the Magnus expansion in the interaction
frame:%
\begin{equation}
I(t)=\int_{0}^{t}dt^{\prime }|\omega _{1}(t^{\prime })|<2\pi  \tag{23}
\end{equation}%
The criterion (23) is the same as the Maricq's criterion (1) and its
derivation is also built on the same scheme as the Maricq's in the case of
single-spin systems. However, it should be pointed out that in the Maricq's
derivation both $\boldsymbol{h(t)}$ and $\boldsymbol{g(t)}$ in Eq. (19) are
the vectors of simple scalars, while in the present situation both the two
vectors $\boldsymbol{h(t)}$ and $\boldsymbol{g(t)}$ are the operator vectors
of the \textit{LOMSO} operator subspace.

\QTP{Body Math}
When a time-dependent shaped RF pulse such as Gaussian pulse [10] and sech
pulse [6] applied to the coupled spin system has a positive amplitude $%
\omega _{1}(t)\geq 0$ over the whole pulse duration $[0$, $t]$ the
integration $I(t)$ equals actually flip angle of the shaped RF pulse. Then
the Magnus expansion in the interaction frame converges and the Magnus
solution of Eq. (9) exists when flip angle of \footnote{%
Here drops a redundant original word.} such shaped pulse is smaller than $%
2\pi ,$ that is, the criterion (23) is met. Evidently, the explicit
criterion (23) serves as a general lower limit for the convergence radius of
the Magnus expansion in the interaction picture [9]. Thus, that the
condition (23) is not fulfilled does not necessarily imply that the Magnus
expansion diverges and the Magnus solution of Eq. (9) does not exist. The
Magnus solution exists really as long as the Magnus criterion is satisfied
[1]. However, the Magnus expansion always converges and hence the Magnus
solution, i.e., the exponential propagator $U_{I}(t)$ of Eq. (9) in the
interaction picture exists certainly when the criterion (23) is met.
Therefore, the exponential propagator of Eq. (9) and the decomposed
propagator of Eq. (11) can be used to determine exactly time evolution of
the coupled spin system S$_{n}$AMX... under a non-negative-amplitude shaped
RF pulse such as Gaussian [10] and sech [6] pulses applied to the spin S
when flip angle of the shaped RF pulse is smaller than $2\pi ,$ a very
popular situation in high-resolution nuclear magnetic resonance spectroscopy.

\QTP{Body Math}
For a weak-amplitude time-dependent RF field the dot product of the two
operator vectors $\boldsymbol{h(t)}$ and $\boldsymbol{g(t)}$ in equation
(19) is usually nearly equal to unit operator. As a consequence, each
eigenvalue of the operator $\hat{\Omega}(t)$ is evaluated approximatively by%
\footnote{%
The approximate formula (24) is not easy to understand. When the amplitude $%
\omega _{1}(t)$ of an RF field is small, the norm $||H_{i}(t)||$ that is
proportional to $|\omega _{1}(t)|$ is also small, as can be seen in (8).
Then in that case the Magnus expansion is quickly convergent [8] and thus
one may use reasonably the lowest-order approximation formula to calculate
the operator $\Omega (t),$%
\[
\Omega (t)\thickapprox \int_{0}^{t}dt^{\prime }H_{i}(t^{\prime }). 
\]%
Now the operator $\Omega (t)$ may be diagonalized by using the unitary
transformation obtained from Eqs. (9) and (11). On the other hand, the
interaction Hamiltonian $H_{i}(t)$ also can be diagonalized by using the
unitary transformation obtained from Eqs. (2) and (6). One therefore obtains
from the lowest-order approximation formula that%
\[
\hat{\Omega}(t)S_{z}\thickapprox \int_{0}^{t}dt^{\prime }\{\omega
_{1}(t^{\prime })V(t,t^{\prime })^{+}S_{z}V(t,t^{\prime })\}. 
\]%
Here $V(t,t^{\prime })$ is a unitary operator. In the trivial case that the
operator $V(t,t^{\prime })$ is the unit operator the formula (24) holds
naturally. In a general case the unitary operator $V(t,t^{\prime })$ may be
written as%
\[
V(t,t^{\prime })=\exp \{-i(V_{x}^{I}(t,t^{\prime
})S_{x}+V_{y}^{I}(t,t^{\prime })S_{y}+V_{z}^{I}(t,t^{\prime })S_{z})\}. 
\]%
Here $V_{p}^{I}(t,t^{\prime })$ $(p=x,y,z)$ are the \textit{LOMSO} operators
of the subsystem AMX.... Note that the amplitude function $\omega _{1}(t)$
and phase function $\phi (t)$ in (2) can be taken as arbitrary functions,
respectively. Then the lowest-order approximation formula shows that one
must set $V_{x}^{I}(t,t^{\prime })\thickapprox 0$ and $V_{y}^{I}(t,t^{\prime
})\thickapprox 0.$ This directly leads the lowest-order approximation
formula to the formula (24).}%
\begin{equation}
\langle i|\hat{\Omega}(t)|i\rangle \thickapprox \int_{0}^{t}dt^{\prime
}\omega _{1}(t^{\prime })  \tag{24}
\end{equation}%
Then the explicit criterion (23) is reduced to the following inequality%
\begin{equation}
\theta (t)=\int_{0}^{t}dt^{\prime }\omega _{1}(t^{\prime })<2\pi  \tag{25}
\end{equation}%
This criterion shows that the Magnus expansion converges and the Magnus
solution (9) exists when flip angle $\theta (t)$ of a weak-amplitude
time-dependent RF pulse is smaller than $2\pi ,$ that is, the inequality
(25) is met. The criterion (25) may be useful for the case of those shaped
RF pulses such as sinc and hermite pulses [4] whose amplitude $\omega
_{1}(t) $ may not be always positive over the whole pulse duration.

\QTP{Body Math}
The real unknown parameters in the propagators of Eqs. (9), (11), and (15)
can be generally determined by solving the linear differential equations
(16). It should be noted that the linear differential equations (16) are not
independent but in which the four operator variables are \textit{constrained}%
\footnote{%
The original manuscript's word is corrected here.} by%
\[
\lbrack \cos (\frac{1}{2}\hat{\Omega}(t))]^{2}+[\cos \alpha (t)\sin \beta
(t)\sin (\frac{1}{2}\hat{\Omega}(t))]^{2} 
\]%
\[
+[\sin \alpha (t)\sin \beta (t)\sin (\frac{1}{2}\hat{\Omega}(t))]^{2}+[\cos
\beta (t)\sin (\frac{1}{2}\hat{\Omega}(t))]^{2}=E 
\]%
where $E$ is unit operator.

\QTP{Body Math}
Actually, the propagator $U_{I}(t)$ of Eq. (5) can be generally expressed as
a sum of the base operators of the operator subspace G%
\begin{equation}
U_{I}(t)=\hat{f}(t)-2i\sum_{p}\hat{g}_{p}(t)S_{p}\text{ \ }(p=x,y,z) 
\tag{26}
\end{equation}%
where the operators $\hat{f}(t)$ and $\hat{g}_{p}(t)$ are the \textit{LOMSO}
operators. Because of the unitarity of the propagator $U_{I}(t)$ the
operators are written respectively in form as $\hat{f}(t)=\cos (\frac{1}{2}%
\tilde{\Omega}(t));$ $\hat{g}_{x}(t)=\cos \tilde{\alpha}(t)\sin \tilde{\beta}%
(t)\sin (\frac{1}{2}\tilde{\Omega}(t)),$ $\hat{g}_{y}(t)=\sin \tilde{\alpha}%
(t)\sin \tilde{\beta}(t)$ $\times \sin (\frac{1}{2}\tilde{\Omega}(t)),$ and $%
\hat{g}_{z}(t)=\cos \tilde{\beta}(t)\sin (\frac{1}{2}\tilde{\Omega}(t)),$
where $\tilde{\alpha}(t),$ $\tilde{\beta}(t),\ $and $\tilde{\Omega}(t)$ are
still the \textit{LOMSO} operators. Obviously, $\tilde{\alpha}(t)=\alpha
(t), $ $\tilde{\beta}(t)=\beta (t),\ $and $\tilde{\Omega}(t)=\hat{\Omega}(t)$
when the Magnus solution (9) exists. By inserting the expansion-form
propagator of Eq. (26) into the Schr\"{o}dinger equation (14) one can still
obtain the same linear differential equations as equations (16) but with the
new operator variables $\hat{f}(t)$ and $\hat{g}_{p}(t)$ $(p=x,y,z)$. The
differential equations are entirely equivalent to the Schr\"{o}dinger
equation (14) and can determine an alternative solution for the Magnus
solution (9) to the time-dependent Schr\"{o}dinger equation [9]. Because the
solution to the differential equations always exists one can obtain the
unknown parameters in the propagator of Eq. (26) by solving the differential
equations. Thus, the expansion-form propagator of Eq. (26) always exists and
can be generally used to determine time evolution of the coupled spin system
under an arbitrary time-dependent RF pulse no matter whether the criterion
(23) is met or not. This indicates that the time evolution can be described
in the small operator algebra subspace $G=u(2)\dbigotimes LOMSO$ instead of 
\footnote{%
Here drops a redundant original word.} the whole Liouville operator space of
the system. Thus, this simplifies greatly determination of the time
evolution.

\QTP{Body Math}
In general, the integration $I(t)$ is always more than flip angle $\theta
(t) $ for a shaped pulse. Then before employing the criterion (23) one
usually needs to first calculate explicitly the integration $I(t)$ for
complex shaped selective pulses such as the BURP pulses [11] and the
Gaussian pulse cascades [12, 13] whose amplitude functions $\omega _{1}(t)$
are not always non-negative over the whole pulse duration even if their flip
angles given in advance are smaller than $2\pi .$ The numerical calculation%
\footnote{%
The numerical calculation merely illustrates that some shaped pulses may
satisfy the criterion (23), while some others also may violate that
criterion in the numerical calculation condition.} shows that the criterion
(23) is met for the excitation pulses $G^{4}$ and $Q^{5}$ of the Gaussian
pulse cascades [12, 13] but is violated for the BURP pulses [11] including
E-BURP, U-BURP, I-BURP, RE-BURP pulses and the Gaussian pulse cascades $%
G^{3} $ and $Q^{3}$ [12, 13]. It should be pointed out that the violation of
the criterion (23) for these pure-phase shaped selective pulses does not
imply that the exponential propagator of Eq. (9) can not be used to describe
time evolution of the coupled spin system under these shaped pulses.
However, as an alternative one still can use the expansion-form propagator
of Eq. (26) of the smaller operator subspace $G=u(2)\dbigotimes LOMSO$
instead of that one of the whole Liouville operator space of the coupled
spin system S$_{n}$AMX... to determine generally time evolution of the spin
system under those shaped pulses and shaped decoupling sequences [17] that
do not satisfy the criterion (23).\newline
\newline
\newline
\textbf{Acknowledgement}

\QTP{Body Math}
This work was supported by the NSFC general projects with grant number
19974064. Author thanks the referees for their valuable suggestions. $%
\newline
$\newline
\textbf{References}\newline
[1] W. Magnus, Commun. Pure Appl. Math. 7 (1954) 649\newline
[2] R. R. Ernst, G. Bodenhausen, and A. Wokaun, \textit{Principles of
nuclear magnetic resonance in one and two dimensions} (Oxford University
Press, Oxford, 1987)\newline
[3] U. Haeberlen, \textit{High resolution NMR in solids}, Adv. Magn. Reson.
Suppl. 1 (1976)\newline
[4] W. S. Warren, J. Chem. Phys. 81 (1984) 5437\newline
[5] X. Miao and R. Freeman, J. Magn. Reson. A 119 (1996) 90\newline
[6] X. Miao, J. Chen, and X. A. Mao, Chem. Phys. Lett. 304 (1999) 45$\newline
$[7] M. M. Maricq, Adv. Magn. Reson. 14 (1990) 151\newline
[8] M. M. Maricq, Phys. Rev. B 25 (1982) 6622\newline
[9] M. M. Maricq, J. Chem. Phys. 86 (1987) 5647\newline
[10] C. Bauer, R. Freeman, T. Frenkiel, J. Keeler, and A. J. Shaka, J. Magn.
Reson. 58 (1984) 442\newline
[11] H. Geen and R. Freeman, J. Magn. Reson. 93 (1991) 93\newline
[12] L. Emsley and G. Bodenhausen, Chem. Phys. Lett. 165 (1990) 469\newline
[13] L. Emsley and G. Bodenhausen, J. Magn. Reson. 97 (1992) 135\newline
[14] X. Miao, Molec. Phys. (in press)$\newline
$[15] X. Miao, X. Han, and J. Hu, Sci. China A 36 (1993) 1199\newline
[16] X. Miao and C. Ye, Molec. Phys. 90 (1997) 499\newline
[17] R. Freeman, \textit{Spin Choreography} (Spektrum, Oxford, 1997)\newline
\newline

\end{document}